\documentclass[aps,pra,twocolumn,superscriptaddress,a4paper,noeprint]{revtex4-1}
\usepackage{graphicx}

\usepackage{amsmath}
\usepackage{float}
\usepackage{color}
\usepackage{textcomp}

\pdfoutput=1 

\usepackage{fancyhdr}
\usepackage[colorlinks=true,citecolor=blue,linkcolor=magenta]{hyperref}
\hypersetup{
pdfauthor = {},
colorlinks = true, linkcolor = blue, urlcolor=blue, bookmarksnumbered =  true}

\begin{document}
\title{Cavity-enhanced Raman scattering for \textit{in situ} alignment and characterization of solid-state microcavities}

\author{Daniel Riedel}
\email{daniel.riedel@caltech.edu}
\altaffiliation{Current address: Institute for Quantum Information and Matter, California Institute of Technology, Pasadena, California 91125, USA}
\affiliation{Department of Physics, University of Basel, Klingelbergstrasse 82, Basel CH-4056, Switzerland}
\author{Sigurd Fl\aa gan}
\affiliation{Department of Physics, University of Basel, Klingelbergstrasse 82, Basel CH-4056, Switzerland}
\author{Patrick Maletinsky}
\affiliation{Department of Physics, University of Basel, Klingelbergstrasse 82, Basel CH-4056, Switzerland}
\author{Richard J. Warburton}
\affiliation{Department of Physics, University of Basel, Klingelbergstrasse 82, Basel CH-4056, Switzerland}
\date{\today}

\begin{abstract}
We report cavity-enhanced Raman scattering from a single-crystal diamond membrane embedded in a highly miniaturized fully-tunable Fabry-P\'{e}rot cavity. The Raman intensity is enhanced 58.8-fold compared to the corresponding confocal measurement. The strong signal amplification results from the Purcell effect. We show that the cavity-enhanced Raman scattering can be harnessed as a narrowband, high-intensity, internal light-source. The Raman process can be triggered in a simple way by using an optical excitation frequency outside the cavity stopband and is independent of the lateral positioning of the cavity mode with respect to the diamond membrane. The strong Raman signal emerging from the cavity output facilitates \textit{in situ} mode-matching of the cavity mode to single-mode collection optics; it also represents a simple way of measuring the dispersion and spatial intensity-profile of the cavity modes. The optimization of the cavity performance via the strong Raman process is extremely helpful in achieving efficient cavity-outcoupling of the relatively weak emission of single color-centers such as nitrogen-vacancy centers in diamond or rare-earth ions in crystalline hosts with low emitter density.
\end{abstract}

\maketitle
\section{Introduction}
The development of a quantum internet crucially relies on the scalable long-distance interconnection of quantum nodes\,\cite{Kimble2008}. These nodes need to combine a robust storage of quantum states and high-fidelity processing of quantum information with an efficient interface to photons mediating the network links via entanglement swapping\,\cite{Wehner2018}.
In order to achieve high entanglement rates these photons need to exhibit transform-limited spectral linewidths, a high degree of single-photon purity and a large creation probability per laser excitation pulse. 
The nitrogen-vacancy (NV) center in diamond constitutes a promising candidate for the stationary qubit due to its highly coherent, optically addressable electron spin along with coupling to multi-qubit nuclear spins in the immediate environment\,\cite{Abobeih2018}.
In seminal proof-of-principle experiments, long-distance entanglement\,\cite{Hensen2015} and on-demand entanglement delivery\,\cite{Humphreys2018} between spatially separated NV centers were demonstrated. However, the entanglement rates are limited to tens of Hertz due to the small fraction ($\sim\,3\%$) of coherent photons emitted into the zero-phonon line (ZPL)\,\cite{Riedel2017}. A promising strategy to overcome this limitation is to enhance the ZPL photon flux of NV centers via coupling to a resonant microcavity\,\cite{Faraon2011,Faraon2012,Li2014,Riedel2017}.

In recent years, tunable Fabry-P\'{e}rot microcavities have been widely used to enhance the photon emission rate of various single emitters\,\cite{Muller2009,Barbour2011,Benedikter2017,Albrecht2013,Greuter2015,Kaupp2016,Johnson2015,Wang2017a,Wang2019,Riedel2017}. The tunability of their resonance frequency in combination with a precise lateral positioning capability allows the emitter-cavity coupling to be maximized \textit{in situ}. A further advantage of this system is that micrometer-scale single-crystalline host materials can be integrated into the cavity while maintaining a high quality factor to mode volume ratio ($Q$/$V$)\,\cite{Janitz2015,Riedel2017}. For emitters which are highly sensitive to fluctuations of the charge environment, increasing the dimensions of a defect-free crystalline environment is clearly beneficial\,\cite{Ruf2019}. An example of such an emitter is the NV center in diamond. NV centers coupled to monolithic nanophotonic structures suffer from spectral fluctuations\,\cite{Faraon2012}; optical performance is better in a Fabry-P\'{e}rot microcavity\,\cite{Riedel2017}.

To maximize their performance, tunable Fabry-P\'{e}rot microcavities require precise \textit{in situ} mode-matching of the cavity mode to external fields. This is in principle simple for a well constructed fiber mirror for which a concave mirror is fabricated at the exact center of an optical fiber\,\cite{Steinmetz2006,Hunger2012}. However, this approach works well only when the mode-field diameter of the cavity is matched to that of the optical fiber. Furthermore, the mode-matching efficiency is inevitably limited by the different wavefront curvatures of the cavity mode and the fiber. Both mismatches are exacerbated for small cavity-mode volumes which require small mirror radii and mode-field diameters. Instead, the ``top" mirror can be fabricated into a silica substrate\,\cite{Barbour2011,Greuter2014,Greuter2015,Riedel2017,Najer2019}, and mode-matching between the cavity and a single-mode fiber can be achieved with a pair of lenses. In practice, this is a non-trivial task. Mode-matching is particularly difficult if two wavelengths are involved, for instance excitation at 532\,nm and NV ZPL emission at 637\,nm, on account of chromatic aberrations. In these experiments, it is also desirable to measure the dispersion of the cavity modes (dependence of the resonance frequency on mirror separation) and the electric field distribution of each mode. These are laborious tasks if a single emitter is used.

We propose here that Raman scattering from the solid-state host is a valuable resource in aligning and optimizing tunable Fabry-P\'{e}rot microcavities and in characterizing the cavity modes. The Raman scattering is enhanced by the cavity and gives large signals, facilitating quick optimization, and subsequently a simple way to determine the mode's dispersion and lateral intensity-profile. The Raman scattering depends at most weakly on the lateral position unlike a single emitter which benefits from cavity enhancement only once the emitter is located at the cavity antinode. 

We report here experiments on diamond from which Raman scattering is well known and has been exploited for both quantum and photonic applications.
Correlated Stokes-anti-Stokes scattering in diamond\,\cite{Kasperczyk2015} led to the development of a macroscopic phonon-based quantum memory\,\cite{Lee2012, England2013, England2015,Anderson2018}, the remote entanglement of macroscopic diamonds\,\cite{Lee2011a} and the development of a Raman laser in the visible wavelength regime\,\cite{Greentree2010, Spence2010}. On account of the large Raman shift of diamond (1{,}332\,cm$^{-1}$\,\cite{Zaitsev2010}) and the high Raman gain coefficient ($\sim$75\,GW$\cdot$cm$^{-1}$ at 532\,nm)\,\cite{Mildren2013}, Stokes scattering provides an excellent narrow-linewidth, high-intensity internal light-source. We show that cavity-enhanced Raman scattering enables fast \textit{in situ} alignment of the diamond cavity-mode with respect to external optics, a fast way of determining the dispersion of the cavity modes, and single-shot imaging of the modes' lateral profile. Additionally, a comparison of the signal with cavity-enhancement to the signal without the cavity is also an indicator of the single-emitter Purcell factor\,\cite{Checoury2010a}. These attributes, demonstrated here on diamond, should be generic to single-crystal solid-state hosts.

\begin{figure}[htb]
\includegraphics[width=0.47\textwidth]{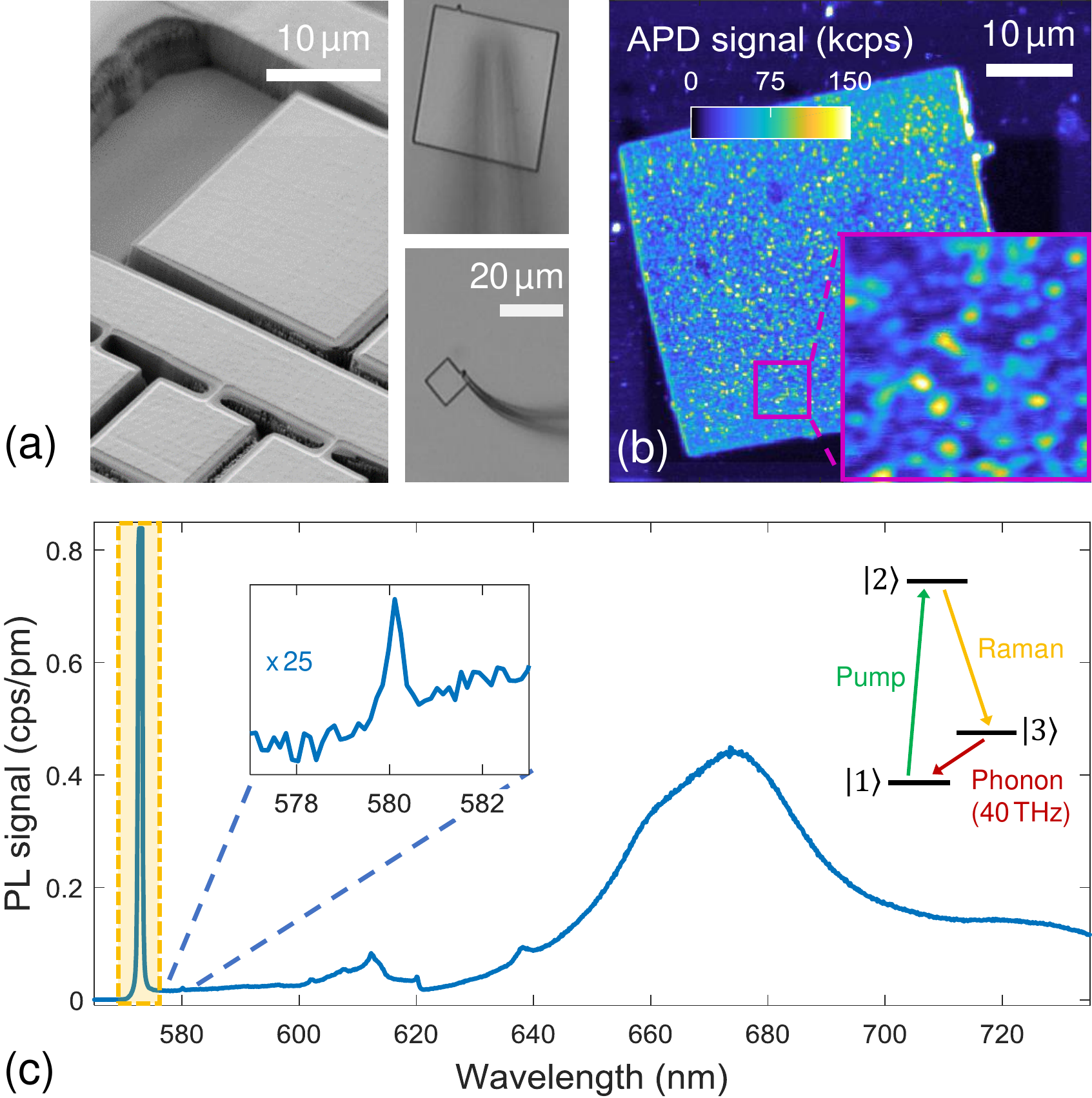}
\caption{(a) Fabrication and transfer of diamond membranes. \textit{Left:} Patterning of the diamond using electron-beam lithography and inductively-coupled plasma-etching. \textit{Right:} Transfer of a membrane using a micromanipulator. (b) Confocal scan of a diamond membrane under green excitation (Laser Quantum Ventus532, $\lambda=532\,$nm, $P=580\,$\textmu W) at room temperature. The emission from single NV centers can be observed. (c) Photoluminescence (PL) spectrum of a single NV center at room temperature. In addition to PL, Raman features are observed: signatures of disordered carbon  and first- and second-order Raman lines of the diamond lattice ($P=3.1\,$mW, integration time 600\,s, for more details see text). \textit{Inset:} Schematic of the Stokes process. A pump photon is converted into a red-shifted Stokes photon and a phonon of a fixed frequency.}
\label{fig:schematic}
\end{figure}

\section{Methods}
We create thin diamond membranes out of high-purity, single-crystal diamond (Element 6) following previously reported fabrication techniques\,\cite{Maletinsky2012, Appel2016}. Using inductively-coupled plasma-etching and electron-beam lithography, we fabricate square-shaped membranes with a typical thickness of $1\,\text{\textmu m}$ and side lengths of $10\,...\,50\,\text{\textmu m}$  (Fig.\,\ref{fig:schematic}(a))\,\cite{Riedel2014, Riedel2017}. The membranes are bonded to a planar SiO$_2$ substrate coated with a highly reflective distributed Bragg-reflector (DBR, 15 layers SiO$_2$/Ta$_2$O$_5$, ECI evapcoat) using a micromanipulator. The extremely smooth surfaces of the diamond membrane (surface roughness $\lesssim0.3\,$nm) and the DBR surface promote strong adherence due to van der Waals forces. The strong bonding is demonstrated by the possibility of bending the micromanipulator needles on attempting to displace the membrane laterally (Fig.\,\ref{fig:schematic}(a)). The bonded membranes contain NV centers which were introduced prior to nano-fabrication by nitrogen-ion implantation and subsequent annealing\,\cite{Riedel2017}. 

As a first step, we characterize the diamond membrane with a room-temperature confocal microscope (i.e.\ without a cavity) using an objective of high numerical aperture ($\textrm{NA}=0.9$). A confocal scan under continuous-wave green excitation ($\lambda=532\,$nm, $P=580\,$\textmu W) exhibits well-isolated bright features which we associate with individual NV centers (Fig.\,\ref{fig:schematic}(b))\,\cite{Doherty2013}.
Fig.\,\ref{fig:schematic}(c) displays a photoluminescence (PL) spectrum for a strong excitation power (532\,nm, 3.1\,mW) and long integration-time (600\,s) recorded at one of these bright spots. The spectrum shows a temperature-broadened NV ZPL at $\sim 637$\,nm and a broad phonon-sideband whose spectral shape is slightly altered in our experiment due to the varying DBR reflectivity with wavelength and thin-film interference in the membrane. Crucially, the spectrum contains clear Raman features: the first- and second-order Stokes features at 572.67\,nm and 600 ... 620\,nm, respectively\,\cite{Zaitsev2010}.
In addition, we find a Raman signature of carbon $\textit{sp}^2$ bonds (Raman shift $\sim$1{,}560\,cm$^{-1}$)\,\cite{Ferrari2000}, indicating either a slight graphitic surface-contamination (which could have been created during high-temperature annealing) or the presence of organic residue.

\begin{figure}[tb]
\includegraphics[width=0.47\textwidth]{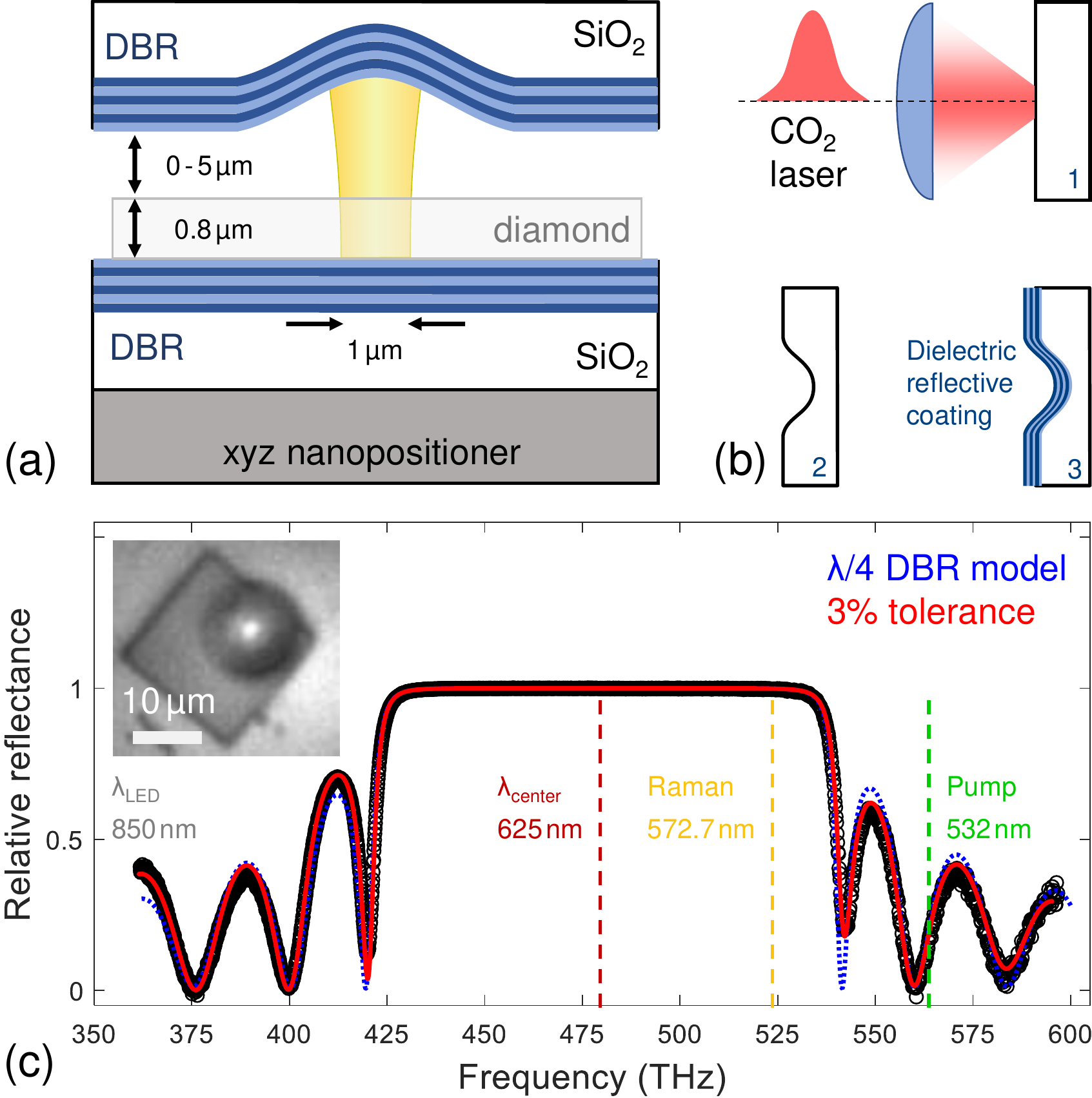}
\caption{(a) Schematic of a diamond membrane embedded in a tunable Fabry-P\'{e}rot microcavity. Nanopositioners enable \textit{in situ} control of both the resonance frequency and the antinode location of the microcavity mode. (b) Process flow of the curved micro-template fabrication. A focused CO$_2$ laser pulse creates a microindentation in a SiO$_2$ substrate via ablation. Subsequently, the template is coated with a dielectric Bragg-reflector (DBR). (c) Normalized white-light transmission spectrum of the DBR coating revealing the mirror stopband. The experiment is well reproduced by a model calculation. Inset: an image of the square-shaped membrane inside the cavity with the microindentation located on top. The bright feature in the center stems from the reflection of laser excitation at wavelength $\sim 635$\,nm.}
\label{fig:cavity}
\end{figure}

Here, we focus on the first-order Stokes-scattering. This process can be modeled by a three-level atom-like system (inset Fig.\,\ref{fig:schematic}(c)) involving a ground state (${\left|1\right>}$) a virtual excited-state (${\left|2\right>}$) and a metastable state (${\left|3\right>}$)\,\cite{Greentree2010}. When ground-state population is excited to state ${\left|2\right>}$, it can de-excite via state ${\left|3\right>}$ by emitting a red-shifted photon and an optical phonon of fixed energy. In our experiment, we determine a spectral shift of $\Delta E=hc \cdot 1{,}335$\,cm$^{-1}$ between the pump laser and the Stokes emission: this corresponds to the optical phonon energy in diamond.

A schematic of our tunable microcavity is shown in Fig.\,\ref{fig:cavity}(a). A planar DBR supporting a diamond membrane ($\sim20\times20\times0.8\,\text{\textmu m}^3$) forms a cavity with a curved DBR. We fabricate an array of atomically-smooth curved microtemplates on a SiO$_2$ chip via CO$_2$-laser ablation yielding small radii of curvature ($R \sim 10\,$\textmu m, Fig.\,\ref{fig:cavity}(b))\,\cite{Hunger2012,Najer2017}. Subsequently, the templates are coated with a highly reflective 14-layer Ta$_2$O$_5$\,/\,SiO$_2$ DBR. The spacing between the two mirrors can be adjusted by applying a voltage to the z-nanopositioner beneath the bottom mirror; the lateral location of the cavity's antinode can be adjusted by applying a voltage to the x- and y-nanopositioners\,\cite{Barbour2011,Greuter2014,Greuter2015,Riedel2017,Najer2019}.

To characterize the mirrors, we measure the transmission spectrum of the planar mirror with a white-light source and quantize the data using the transmission spectrum of an uncoated quartz substrate (Fig.\,\ref{fig:cavity}(c)). With a transfer-matrix calculation we are able to reproduce the oscillations of the reflectivity over a large frequency range. A transfer matrix-based refinement algorithm allows the reflection spectrum to be reconstructed on setting an individual layer-thickness tolerance of 3\,\% (Essential MacLeod). For our calculation we set a stopband center of $\lambda_{\text{center}}=625\,$nm and use 15 $\lambda$/4 layer pairs of SiO$_2$ and Ta$_2$O$_5$ with a refractive index of n$_\text{SiO$_2$}=1.46$ and of n$_\text{Ta$_2$O$_5$}=2.11$, respectively. For the top mirror we obtain similar results, reproducing the transmission spectrum with $\lambda_{\text{center}}=629\,$nm and 14 layer pairs.

To cavity-enhance the Raman process, we pump the diamond with a green laser and tune the cavity into resonance with the Stokes line. The pump laser can be coupled into the cavity independently of the mirror separation since its wavelength (532\,nm) lies outside the reflection stopband of the mirror coating (Fig.\,\ref{fig:cavity}(c)). Conversely, due to the large Raman shift of diamond, the cavity supports a resonance with a finesse of $\sim 1{,}000$ at the Stokes wavelength $\sim 573$\,nm. 

We mention some of the details of the experiment. An infrared light-emitting-diode (LED, $\lambda_{\text{LED}}=850$\,nm for which the DBR is transparent) is used to locate the membrane inside the cavity (inset Fig.\,\ref{fig:cavity}(c)). This visual feedback allows us to adjust the location of the diamond with respect to the top mirror. The pump laser is spectrally filtered (Semrock, LL01-532-25 and FF01-650/SP-25) and coupled into the cavity via a slightly overfilled objective of moderate numerical aperture (Microthek, 20x/0.4). The Stokes signal is collected via the same objective and coupled into a single-mode fiber (Thorlabs 630-HP); the output of this fiber is coupled into a spectrometer. A combination of a dichroic mirror (Semrock, FF560-FDi01) and longpass/bandpass filters (Semrock, LP03-532RS-25 and FF01-572/15-25) is employed to prevent laser light entering the collection optics. This is important to avoid exciting fluorescence and Raman scattering from the detection fiber. Using a precision mechanical-stage we can move the entire microcavity with respect to the external optics allowing the cavity output to be aligned with respect to the optical axis of the microscope.

\begin{figure}[t]
\includegraphics[width=0.47\textwidth]{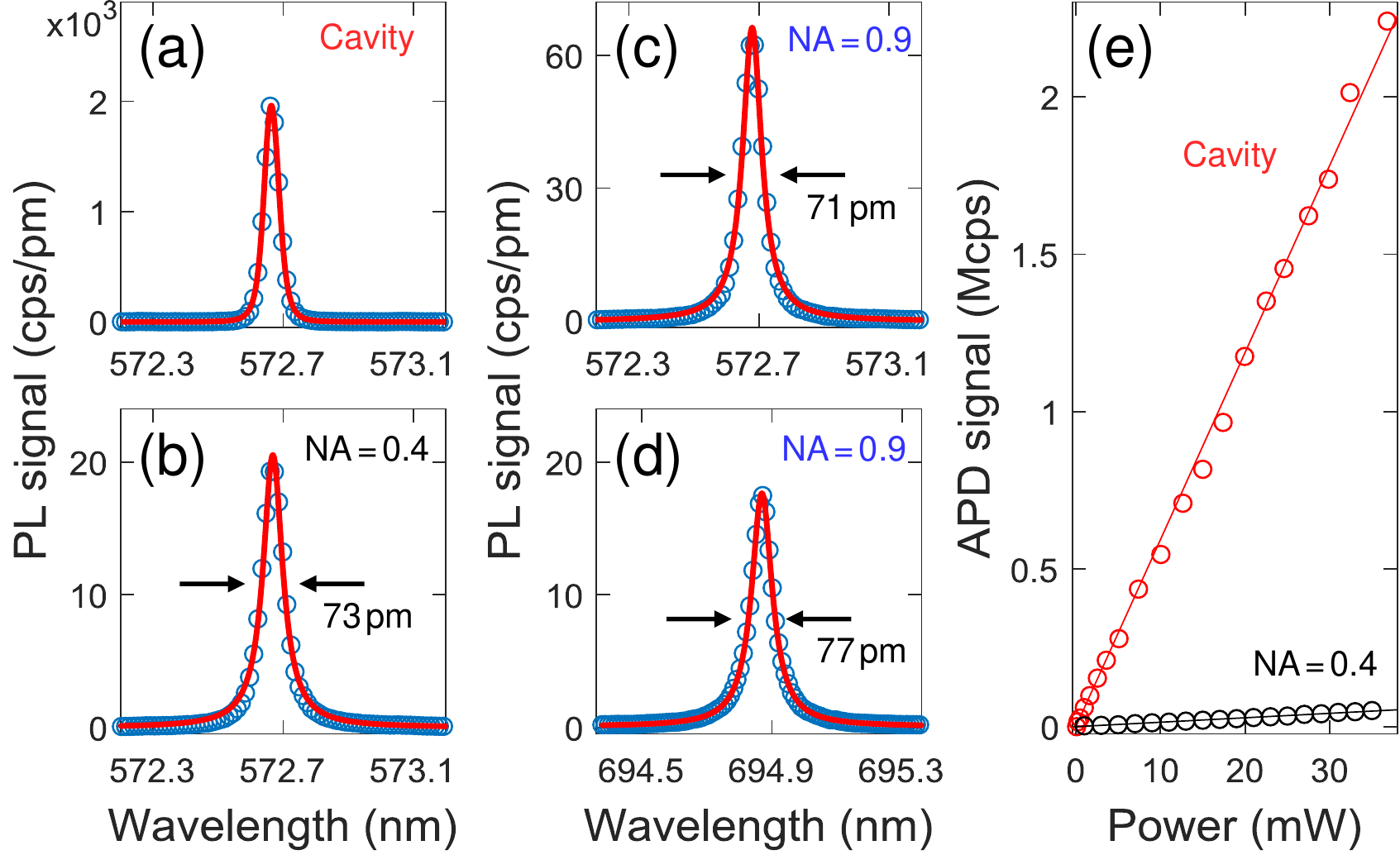}
\caption{(a) Typical optical spectrum of the Stokes signal in resonance with the cavity ($t_{\text{int}}=1$\,s). (b) Typical optical spectrum of the Stokes signal for out-of-cavity detection using an objective with NA=0.4 ($t_{\text{int}}=120$\,s), and (c) NA=0.9 ($t_{\text{int}}=30$\,s). (d) Typical optical spectrum of the Stokes signal pumped with a narrow-band laser at $\lambda=636$\,nm ($P=5$\,mW) without the top mirror using an objective of NA=0.9 ($t_{\text{int}}=180$\,s). (e) Power dependence of the integrated Stokes signal for out-of-cavity detection with an objective of NA=0.4 along with the integrated cavity-enhanced Stokes signal. (a), (b) and (c): The pump wavelength is $\lambda=532$\,nm, the pump power $P=20\,$mW. The Stokes lineshape slightly deviates from a Lorentzian due to the spectral profile of the pump laser; all data were recorded at room temperature.}
\label{fig:power}
\end{figure}

\section{Results}
Figure\,\ref{fig:power}(a) shows a spectrally resolved measurement of the cavity-enhanced Stokes emission at $\lambda_\text{s}=572.67\,$nm pumped at 532\,nm with $P=20$\,mW. Attaining the maximal signal strength requires careful alignment. This process is massively aided by exploiting the Raman process on account of at least three factors. First, the signal is very large. When the cavity is tuned into resonance with the Stokes photons, we detect up to  $2 \times 10^{6}$ photon counts/s (cps) on a standard silicon single-photon avalanche-photodiode. Secondly, the signal does not depend on the $(x,y)$-alignment of the cavity: it represents a ubiquitous internal light-source. Thirdly, the Raman process couples to all the cavity modes. From the mode dispersion and the signal strengths, this allows the transverse electromagnetic (TEM) cavity mode indices $\text{(q,n,m)}$ to be determined. In particular, the $\text{(q,0,0)}$ modes can be identified: it is these modes whose output couples best to the single-mode fiber detection-channel. 


The photon flux of the cavity-enhanced Stokes scattering is strongly enhanced with respect to the Stokes signal collected from the bare diamond membrane under equivalent experimental conditions (Fig.\,\ref{fig:power}(a)). Integrating over wavelength in both cases (with cavity, without cavity), we find an enhancement factor of $F=58.8$. This enhancement results from the Purcell effect\,\cite{Cairo1993}. At a given power, the cavity increases the Stokes photon generation rate by the Purcell factor, $F_\text{P}$. Additionally, the Stokes photons are emitted preferentially into the cavity mode -- this enhances the detection efficiency by a factor $F_\text{c}$. Overall, $F=F_\text{P}\cdot F_\text{c}=58.8$. The Stokes signal increases linearly with pump power (Fig.\,\ref{fig:power}(e)): at these pump powers, there is no super-linear dependence presaging Raman lasing.


\begin{figure*}[bt]
\includegraphics[width=17.8cm]{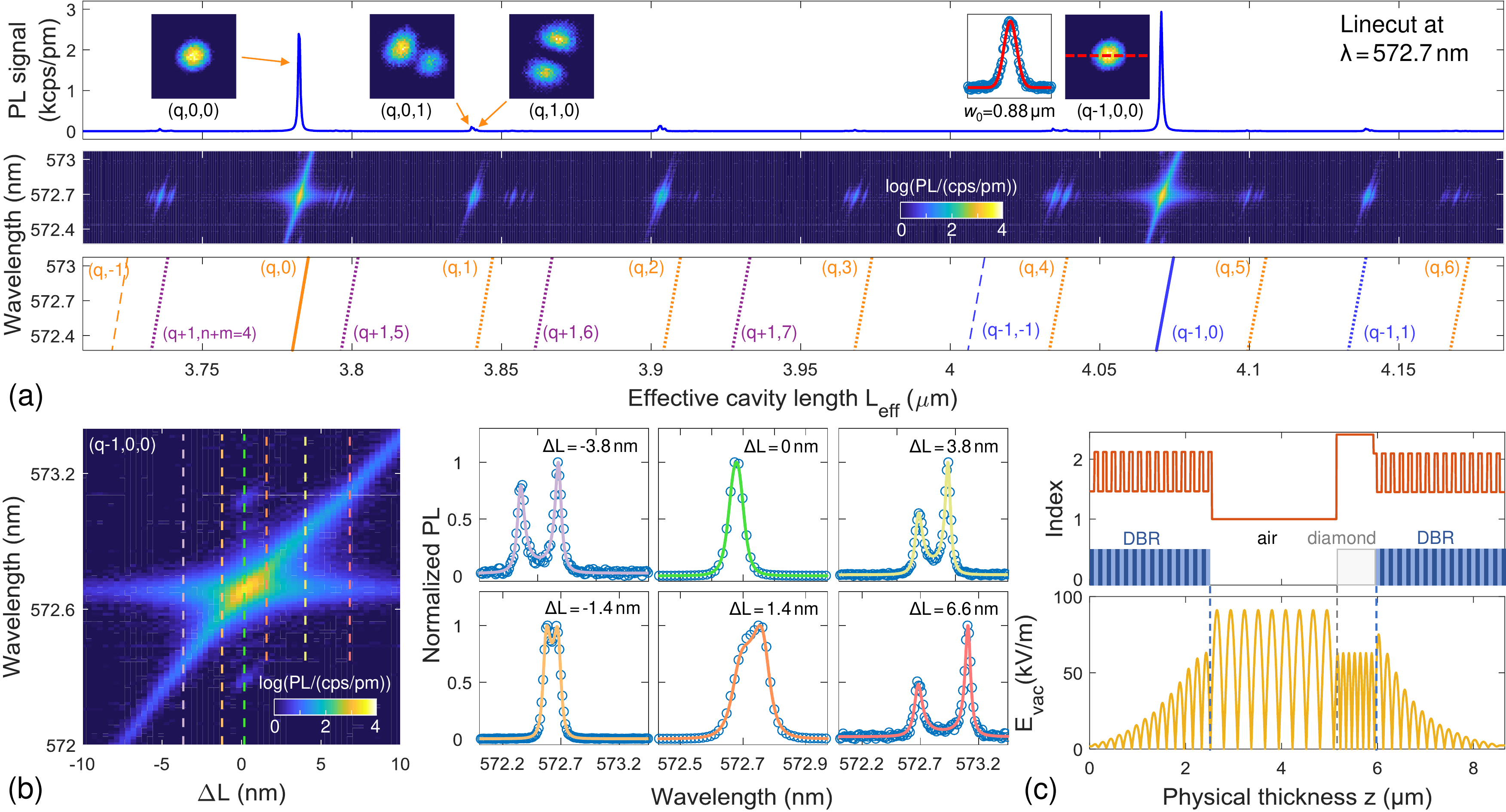}
\caption{(a) \textit{Center panel:} PL spectra for different relative cavity lengths $\Delta L$ about $L=4.07\,$\textmu m under green illumination ($\lambda=532\,$nm, $P=20\,$\textmu W) detected with a single-mode fiber.
\textit{Top panel:} Linecut at $\lambda=572.67$\,nm. The Stokes light couples to the Gaussian cavity-modes (q,n,m) of different mode families. Well isolated (q,0,0), (q,1,0) and (q,0,1) modes can be directly imaged on a CCD camera. \textit{Bottom panel:} Calculation of the mode dispersion using an analytic model based on Gaussian optics. The spacing of the different modes allows the geometric parameters of the cavity to be extracted (for details see text).
(b) Zoom-in of the spectra in (a) featuring linecuts at different $\Delta L$ (for mode (q-1,0,0)). The data can be fitted well using two multiplied Lorentzians (for details see text). Data in (a) and (b) were recorded at room temperature. (c) A one-dimensional transfer-matrix simulation of the electric vacuum field for $t_\text{a}=2{,}596$\,nm and $t_\text{d}=772$\,nm.}
\label{fig:modes}
\end{figure*}



We now exploit the Raman process as a convenient way to analyze the cavity modes. To do this, we increase gradually the cavity length, monitoring the cavity emission at the Stokes wavelength. Fig.\,\ref{fig:modes}(a) displays the spectrally resolved measurement of the cavity emission (Fig.\,\ref{fig:power}(b)) for different cavity length detunings $\Delta L$ recorded by adjusting the width of the air-gap $t_\text{a}$. A series of cavity modes is observed. In this experiment, the linewidth of the fundamental mode $\delta t_\text{a}$ is a measure of the cavity finesse $\mathcal{F}=(\lambda/2)/\delta t_\text{a}=350$ (Fig.\,\ref{fig:modes}(a) top panel)\,\cite{Greuter2014}. Clearly, the Raman process couples to the various higher-order Gaussian modes whenever a spectral resonance with the Stokes photons is established. The collection efficiency of the higher-order modes is much lower than that of the fundamental modes due to the signal collection through a single-mode fiber but nevertheless a number of higher-order modes are observed.

The exact locations of the cavity modes depend on the cavity geometry: an analysis of the spacings of the different modes allows the geometric parameters of the cavity to be determined. We extract the radius of curvature of the top mirror, $R_\text{cav}$, from the spacings between the fundamental mode (q,0,0) and its associated higher harmonics (q,n,m). Quantitatively, we make this link with a Gaussian optics model\,\cite{Greuter2014}. The effective cavity length $L_\text{eff}$, $R_\text{cav}$ and the (q,n,m)-parameters are connected by
\begin{equation}
 L_\text{eff}({\text q,n,m})=\left[\text{q}+\frac{\text{n}+\text{m}+1}{\pi}\arccos{(\sqrt{g})}\right]\cdot\frac{\lambda}{2}
\end{equation}
where $g$, the confocal parameter, is given by $g=1- L_\text{eff}({\text q,n,m})/R_\text{cav}$. $L_\text{eff}$ is a measure of the separation of the two mirrors accounting for the penetration depth into the mirrors upon reflection. From this model, we find that the modes in Fig.\,\ref{fig:modes}(a) are well described with $R_\text{cav}=10\,$\textmu m (Fig.\,\ref{fig:modes}(a) lower panel).

The signals in these experiments are sufficiently large that the spatial intensity-distribution of the modes can be recorded in a single-shot imaging experiment. We directly image the modal shape of the fundamental \text{(q,0,0)} and the first two higher-order modes \text{(q,1,0)} and \text{(q,0,1)} on a charge-coupled-device camera (Fig.\,\ref{fig:modes}(a)).
Using the diamond membrane (edge length 20\,\textmu m) as a ruler, we can calibrate the lateral dimensions of the images. A Gaussian fit of a linecut through the fundamental mode yields a beam waist of $w_\text{I}=0.88\,$\textmu m (Fig.\,\ref{fig:modes}(a), inset). This value corresponds to the beam waist at the top mirror ${w_\text{I}(z=4.07\,\text{\textmu m},R(z)=10\,\text{\textmu m})=0.87\,\text{\textmu m}}$ calculated from Gaussian optics.

The separation of the fundamental (q,0,0) resonances, specifically the change in resonance wavelength per change in air-gap width $m=\Delta \lambda_\text{c}/\Delta t_\text{a}$, allows the effective cavity mode number $q$ to be inferred. For the two (q,0,0) resonances in Fig.\,\ref{fig:modes}(a) we measure $m_1=87\,\text{pm}/\text{nm}$ and $m_2=83\,\text{pm}/\text{nm}$ corresponding to $q_1=2/m_1=23$ and $q_2=2/m_2=24$. The full cavity mode-structure must be modeled including all the interferences\,\cite{Barbour2011}. Conceptually, the cavity modes can be described using a coupled-cavity approach\,\cite{Janitz2015,Riedel2017,VanDam2018}. In this picture, there are two cavity modes, one defined by the air-gap bounded by the top DBR and the diamond-air interface; the other is defined by the diamond-air interface and the bottom DBR layer (Fig.\,\ref{fig:modes}(c)). The two modes couple via the non-zero reflectivity of the diamond-air interface and hybridize.

We simulate the cavity modes with the aid of the same software we used to reconstruct the stopband of the DBR mirrors (Fig.\,\ref{fig:cavity}(c)). The exact mirror structure is included. $q_1$ tells us that the lowest fundamental mode is the 8th resonance (5th resonance away from contact; curved mirror depth $>3 \lambda/2$). The gradients $d \lambda_\text{c}/d t_\text{a}$ of the (q,0,0) modes depend on the exact diamond thickness $t_\text{d}$ and the exact air-gap thickness $t_\text{a}$. By adjusting $t_\text{d}$ and $t_\text{a}$ in the simulation, we match the experimental results for $m_1$ and $m_2$ with $t_\text{d}=0.77\,$\textmu m and $t_\text{a}=2.60\,$\textmu m.
The diamond thickness is in agreement with the value we found in our previous NV coupling experiment\,\cite{Riedel2017} which used the exact same membrane.

We now turn to analyzing the \text{(24,0,0)} mode in more detail. This particular mode is well-isolated and not perturbed by coupling to higher-order modes of other mode families (Fig.\,\ref{fig:modes}(b))\,\cite{Benedikter2015}. In particular, we focus on the spectral properties as the cavity is tuned into resonance with the Stokes process. 
Without the top mirror (i.e.\ no cavity), we find that the Stokes resonance has a Lorentzian lineshape with a full width at half maximum (FWHM) of $\delta \lambda_\text{s,532}=71\,$pm (Fig.\,\ref{fig:power}(b)). This linewidth, 64.9\,GHz, is determined by a convolution of the laser linewidth, $\sim15$\,GHz, with the linewidth of the Stokes scattering process, $\delta\nu_\text{s}\sim50$\,GHz. We measure $\delta\nu_\text{s}$ independently by pumping the Stokes process with a narrow-bandwidth laser at $\lambda=636$\,nm and find $\delta\lambda_\text{s,636}=77\,$pm corresponding to $\delta\nu_\text{s}=47.8$\,GHz (Fig.\,\ref{fig:power}(d)). In the absence of inhomogeneous strain fields, the Raman linewidth is a measure of the phonon lifetime. The measured Raman linewidth corresponds well with previously reported values, 3.6\,...\,3.9\,ps (40.8\,...\,44.2\,GHz)\,\cite{Lee2012,Anderson2018}, indicating low strain in our diamond membrane. Here, the main decay channel of the optical phonon involves the creation of two acoustic phonons each with lower energy\,\cite{Klemens1966, Liu2000}. With the top mirror, we tune the cavity through the Stokes resonance, recording spectra at each detuning (Fig.\,\ref{fig:modes}(b)). The experimental spectra are well fitted by the product of two Lorentzians describing the cavity and the Stokes process, $\mathcal{L}_\text{c}(\lambda_\text{c},\delta \lambda_\text{c})\cdot\mathcal{L}_\text{s}(\lambda_\text{s},\delta \lambda_\text{s}) $.
During the experiment we tune the resonance frequency of the cavity $\lambda_\text{c}$ while $\lambda_\text{s}=572.67\,$nm and $\delta \lambda_\text{s,532}=71\,$pm are fixed.
From our fit we extract that the linewidth of the cavity $\delta \lambda_\text{c}$ decreases from 80\,pm (73 GHz) to 60\,pm (55 GHz) on detuning from wavelength 572\,nm to 573.4\,nm (corresponding to Q factors of 7{,}200 and 9{,}600, respectively) due to the change in mirror reflectivity. (In this experiment, the bare Stokes resonance has a similar spectral width as the cavity resonance.) On resonance $\lambda_\text{c}=\lambda_\text{s}$, $\delta \lambda_\text{c}=70$\,pm and $Q_\text{c,res}=$~8{,}200. The slight deviation of the Stokes lineshape from a Lorentzian (see Fig.\,\ref{fig:power}) results in an $\sim 10\%$ error of the extracted linewidth.

The spectral information leads to an interpretation of the microscopic nature of the cavity-enhanced Raman process. The spectra do not mimic the behavior of a coherent single emitter coupled to a single cavity-mode. Instead, they mimic the behavior of an independent array of emitters (described by $\mathcal{L}_\text{s}$) coupled to a single cavity mode (described by $\mathcal{L}_\text{c}$). The Lorentzian cavity mode $\mathcal{L}_\text{c}$ acts as a spectrally-selective booster for the Raman processes which are resonant with it; the Raman medium consists of a band of independent photon/phonon modes. 

We attempt to understand the enhancement factor $F$ quantitatively\,\cite{Hummer2016}. The first step is to calculate the Purcell factor. One way to describe the signal enhancement promoted by the cavity is to consider an increase of the effective Stokes scattering length. The cavity finesse is a measure of the number of times a photon bounces between the mirrors and corresponds to the factor by which the cavity length is increased. This description is formally equivalent to the Purcell formula for a cavity formed by two mirrors\,\cite{Reiserer2015}.
In our experiment, however, the cavity is more complex than a generic Fabry-P\'{e}rot device. We calculate the vacuum electric-field distribution with the same software we used for the previous calculations of the mirror reflectivity and the slope of the mode dispersion (Essential MacLeod). To that end, we calculate the electric field distribution of a one-dimensional cavity using the geometric parameters extracted from Fig.\,\ref{fig:modes}(a). We then quantize the field amplitude of the Gaussian cavity mode according to:
\begin{equation}
\begin{split}
\int_{\rm cav}\epsilon_0 \epsilon_\text{R}(z) |\vec{E}_\text{vac}(z)|^2 {\rm d}z\int_0^{2\pi}{\rm d}\phi\int_0^{\infty} r e^{-r^2/2 w_\text{I}^2}{\rm d}r =\\
=2 \pi \cdot \frac{1}{4}w_\text{I}^2 \int_{\rm cav}\epsilon_0 n^2(z) |\vec{E}_\text{vac}(z)|^2 {\rm d}z=\hbar \omega /2
\end{split}
\end{equation}
We take $\epsilon_\text{R}=n^2$; we assume a constant beam-waist $w_\text{I}$ over the cavity length. Taking a representative value, $w_\text{0,I}=0.77\,$\textmu m (an average of the beam-waist at the top mirror and the minimum beam-waist), we find a maximum $|\vec{E}_\text{vac}(z)|=54.4$\,kV/m inside the diamond from which we can calculate the effective mode volume of a cubic resonator made from diamond to be $V_\text{eff}=84.9 (\lambda/\text{n})^3$. With the cavity $Q$-factor on resonance $Q_\text{c}(\lambda_\text{c})=$~8{,}200, we calculate the Purcell factor to be ${F_\text{P}(\lambda_\text{c})=1+3/(4 \pi^2)\cdot Q_\text{c}/V\cdot(\lambda/\text{n})^3\cdot1/2=4.7}$. Here, the factor $1/2$ takes into account averaging of the enhancement over the field profile inside the diamond. The second step is to calculate the coupling efficiencies. Without the top mirror, we estimate the coupling efficiency $\eta_\text{o}$ simply as twice the solid-angle defined by the objective lens: $\eta_\text{o} \simeq 1-\sqrt{1-(\text{NA}/n_\text{d})^2}$. (The factor of two accounts for reflection from the bottom DBR.) With the cavity, the coupling efficiency is given by $\eta_\text{c}=\kappa_\text{t}/(\kappa_\text{t}+\kappa_\text{b}) \cdot \beta$ where $\kappa_\text{t}$ ($\kappa_\text{b}$) is the loss rate through the top (bottom) mirror and $\beta=F_\text{P}/(F_\text{P}+1)$ is the probability of emission into the cavity mode. Assuming an identical pump rate, likewise identical diamond material parameters and collection optics, the ratio of the Stokes signal with cavity enhancement to the Stokes signal without the cavity (i.e.\ no top mirror) is $S_\text{c}/S_\text{o}=F_\text{P} \cdot \eta_\text{c}/\eta_\text{o}$. A final factor is that the Stokes $(Q_\text{s})$ and cavity linewidths $(Q_\text{c})$ are similar\,\cite{Checoury2010a}: the final result is $S_\text{c}/S_\text{o} \simeq F_\text{P} \cdot Q_\text{s}/(Q_\text{s}+Q_\text{c}) \cdot \eta_\text{c}/\eta_\text{o}$. Quantitatively, this result predicts that $S_\text{c}/S_\text{o}=56.8$, in good agreement with the experiment. (We note that the error in the experimental measurement of $S_\text{c}/S_\text{o}$ is dominated by a systematic error of 10\% arising on ensuring that the optical alignment is preserved on removing the top mirror; the error in the calculated $S_\text{c}/S_\text{o}$ is dominated by the uncertainties of $\sim10\%$ in the beam-waist and Q factor of the cavity.) It is interesting to note that $S_\text{c}/S_\text{o}$ is much simpler to measure than the Purcell factor of a single emitter: a measurement of $S_\text{c}/S_\text{o}$ is an indicator of the performance of a single-emitter in the same cavity structure. 

The mode-structure calculation shows that the cavity is in a ``diamond-like'' configuration at the Stokes wavelength $\lambda_\text{s}$\,\cite{Janitz2015}. 
An advantage of this configuration is that the electric vacuum field is more strongly confined in the diamond layer than in the air-gap which leads to higher coupling strengths. A further advantage is that the dispersion of the mode wavelength with air-gap thickness is relatively small. This renders the cavity less susceptible to acoustic noise. However, a diamond-like cavity exhibits a vacuum field antinode at the diamond-air interface which exacerbates losses (over a configuration with a field node at the diamond-air interface) caused by scattering at this interface or absorption of the surface. In some materials, GaAs for instance, these losses can be mitigated by passivating the surface\,\cite{Guha2017,Najer2019}.

\begin{figure}[tb]
\includegraphics[width=0.47\textwidth]{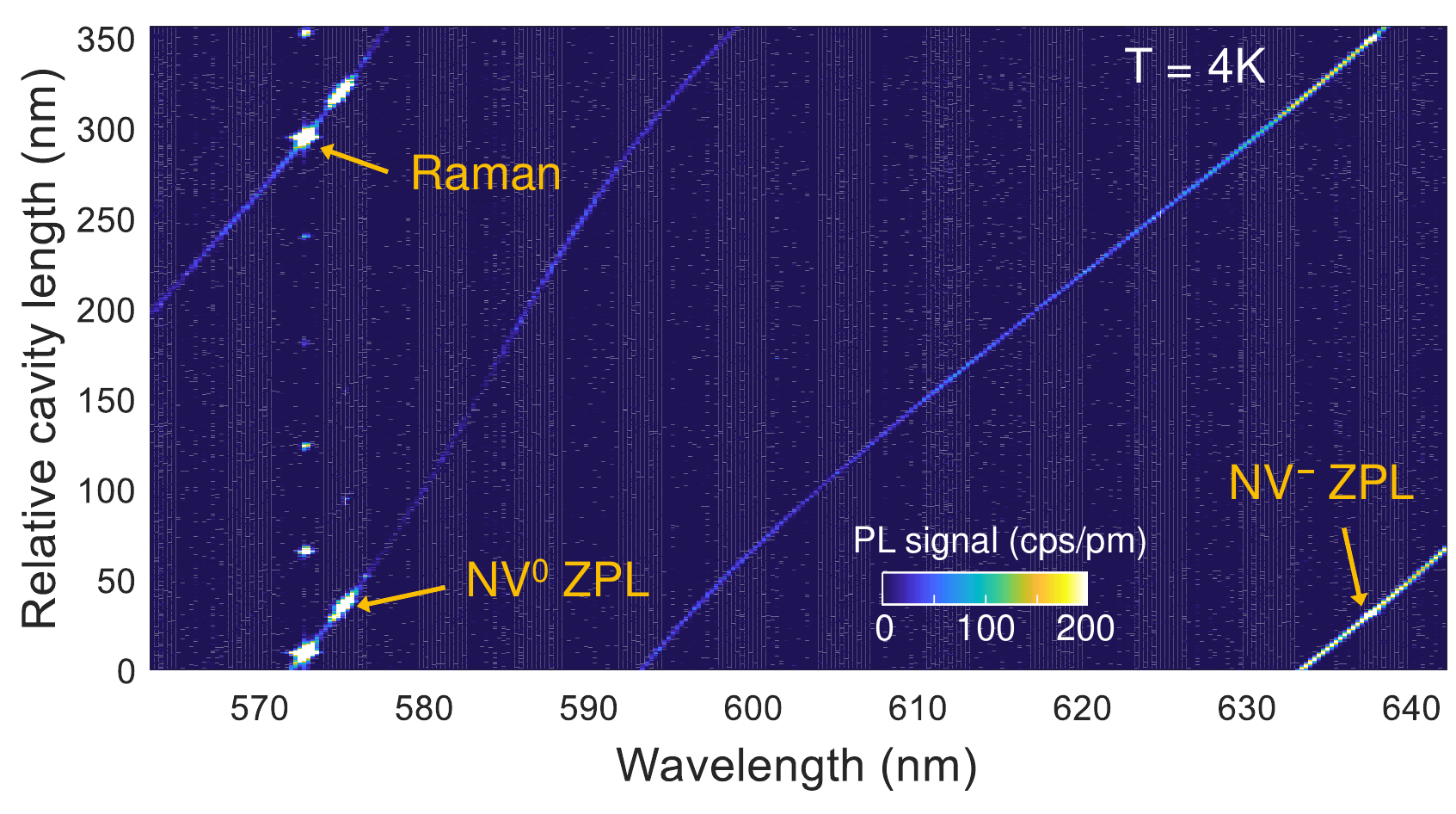}
\caption{Mode structure of the cavity recorded at a temperature of 4\,K, under continuous-wave excitation at 532\,nm with a power of 30\,mW. In addition to the Raman transition, the ZPL of the $\textrm{NV}^0$ and $\textrm{NV}^-$ couples to the cavity.} 
\label{fig:outlook}
\end{figure}

The results presented so far were all recorded at room temperature. For single emitters, operation at low temperature is necessary in order to eliminate phonon-related broadening of the ZPL. We therefore  demonstrate that the cavity-enhanced Raman scattering works well also at cryogenic temperature. The compact cavity design facilitates low-temperature experiments in a liquid-helium bath cryostat\,\cite{Greuter2015,Riedel2017,Najer2019}. Fig.\,\ref{fig:outlook} shows a spectrally resolved cavity measurement where the cavity length is tuned over one free spectral range for an excitation at wavelength 532\,nm. The weak background PL of the diamond allows the main cavity modes to be observed at all cavity lengths and in Fig.\,\ref{fig:outlook}, the nonlinear mode-structure is visible. The first-order Raman peak and the ZPLs of the two different charge states of the NV are enhanced by the cavity and are very strong features in Fig.\,\ref{fig:outlook}. The weak emission of the ZPLs is only detected efficiently once the external optical excitation and collection are properly aligned. The strong Raman signal was used to achieve the alignment: the advantage secured at room temperature therefore translates directly to operation at a cryogenic temperature.
\section{Conclusion}
We show that Raman scattering provides a valuable resource in optimizing and quantifying the performance of tunable micro-cavities. We apply this method specifically to the NV center in diamond. More generally, the generic nature of Raman scattering renders our approach immediately applicable for improving the spin-photon interface efficiencies of a wide range of solid-state qubits. This will be particularly valuable for emitters with a weak oscillator strength: by harnessing the Stokes process as a strong, narrowband, internal light-source, the cavity performance can be optimized, facilitating the detection of signals from weak single-emitters. 
Examples of qubits with long spin coherence times but small optical dipole-moments include color centers in silicon carbide\,\cite{Riedel2012, Christle2017} and rare earth ions\,\cite{Dibos2018, Zhong2018,Casabone2018}. 

\section*{Acknowledgements}
We acknowledge financial support from NCCR QSIT, a competence center funded by SNF, the Swiss Nanoscience Institute (SNI),  ITN network SpinNANO and the EU Quantum Flagship project ASTERIQS (Grant No. 820394).


\bibliographystyle{apsrev4-1}

\end{document}